\documentclass[useAMS,usenatbib,usegraphicx]{mn2e}

\title[Non-thermal emission from molecular clouds]{Broad-band nonthermal emission from molecular clouds illuminated by cosmic rays from nearby supernova remnants}
\author[S. Gabici, F. A. Aharonian, and S. Casanova]{S. Gabici$^{1}$\thanks{E-mail:
sgabici@cp.dias.ie}, F. A.
Aharonian$^{1,}$ $^{2}$, and S. Casanova$^2$ %\footnotemark[1]\thanks{This file has been amended to
\\
$^{1}$Dublin Institute for Advanced Studies, 31 Fitzwilliam Place, Dublin 2, Ireland\\
$^{2}$Max--Planck--Institut f\"ur Kernphysik, Saupfercheckweg 1, 69117 Heidelberg, Germany}
\begin{document}

\date{Accepted . Received ; in original form }

\pagerange{\pageref{firstpage}--\pageref{lastpage}} \pubyear{}

\maketitle

\label{firstpage}

\begin{abstract}
Molecular clouds are expected to emit non-thermal radiation due to cosmic ray interactions in the dense magnetized gas. Such emission is amplified if a cloud is located close to an accelerator of cosmic rays and if energetic particles can leave the accelerator site and diffusively reach the cloud. We consider here the situation in which a molecular cloud is located in the proximity of a supernova remnant which is efficiently accelerating cosmic rays and gradually releasing them in the interstellar medium. We calculate the multiwavelength spectrum from radio to gamma rays which is emerging from the cloud as the result of cosmic ray interactions. The total energy output is dominated by the gamma ray emission, which can exceed the emission in other bands by an order of magnitude or more. This suggests that some of the unidentified TeV sources detected so far, with no obvious or very weak counterparts in other wavelengths, might be in fact associated with clouds illuminated by cosmic rays coming from a nearby source. Moreover, under certain conditions, the gamma ray spectrum exhibit a concave shape, being steep at low energies and hard at high energies. This fact might have important implications for the studies of the spectral compatibility of GeV and TeV gamma ray sources.
\end{abstract}

\begin{keywords}
\end{keywords}

\section{Introduction}

Galactic Cosmic Rays (CRs) are believed to be accelerated via first order Fermi mechanism at the expanding shock waves of supernova remnants (SNRs) \citep[for a review see][]{drury,blandfordeichler,jonesellison}.
Two of the strongest points supporting this scenario are: \textit{i)} the fact that SNRs alone are capable to maintain the galactic CR flux at the observed level, provided that a fraction of about 10\% of their explosion energy is somehow converted into CRs \citep{baade,hillas}, and \textit{ii)} the fact that some shell-type SNRs have been detected at TeV energies \citep[see][for a review]{felixrev}, as expected if they indeed are the sources of CRs \citep{dav1}.
In this context, TeV gamma rays originate from the decay of neutral pions produced in interactions between the accelerated CRs and the interstellar gas swept up by the SNR shock.

However, the detection of SNRs in gamma rays, though encouraging, cannot provide by itself the final proof that CRs are indeed accelerated at SNR shocks. This is because competing leptonic processes, namely inverse Compton scattering from accelerated electrons, can also explain the observed TeV emission, provided that the magnetic field does not significantly exceed $\sim 10 ~ \mu$G (e.g. \citealt{hessrxjlong}). Evidence for strong $\approx 100 ~ \mu$G magnetic field, and thus indirect support to the hadronic scenario for the gamma ray emission, comes from the observation of thin X-ray synchrotron filaments surrounding some SNRs \citep{bamba,vink,heinz} and of the rapid variability time scale of the synchrotron X--rays \citep{ias1,ias2}.
On the other hand, the absence of thermal X-ray emission from the TeV SNRs RXJ1713.7-3946 and RXJ0852.0-4622 has been interpreted by Katz and Waxman (2008) as an argument against the hadronic nature of the gamma-ray emission (and thus supporting the leptonic scenario).
Their argument can be summarized as follows: in order to fit the TeV data with an hadronic model, the gas density cannot be too low. The high density would then imply a strong X-ray thermal Bremsstrahlung emission which is not observed.
However, this argument is not robust given the uncertainties related to the heating of electrons at shocks, and it has been strongly questioned \citep{tanaka,morlino,volklast,mereview,luke}. 
Moreover, it has been shown that the downstream proton temperature can be effectively reduced if the shock is effectively accelerating cosmic rays. This would also suppress the electron temperature and the thermal X-ray emission \citep{luke,jacco}.
A conclusive proof of the hadronic nature of the gamma-ray emission will possibly come from the detection of neutrinos from SNRs, which are expected to be produced during the same hadronic interactions responsible for the production of gamma rays \citep{kappes,vissani,gabici2}.
However, since the detection of neutrinos appears challenging even for km$^3$-scale neutrino telescopes, the search for evidence of CR proton acceleration coming from gamma ray or multi wavelength observations is mandatory. 

The presence of a massive molecular cloud close to a SNR can provide a dense target for CR hadronic interactions and thus enhance the expected gamma-ray emission. A correlation between SNRs and molecular clouds is expected, especially in star forming regions \citep{snobs}. In general, a spatial correlation between TeV gamma rays and dense gas, though not conclusive, would favor an hadronic scenario, where gamma rays are the result of the interactions of CR protons in the dense gas. Such correlation has been observed from a number of SNR/molecular clouds associations \citep{hess1745,ic443,w28}.
If the molecular cloud is overtaken by the SNR shock, the enhanced gamma-ray emission is expected to be cospatial with the SNR shell, or with a portion of it \citep{dav2}. If the cloud is located at some distance from the SNR, it can still be illuminated by CRs that escape from the SNR and produce gamma rays there \citep{atoyan,gabici2}. For this scenario, it has been shown that, for typical SNR parameters and for a distance $D = 1$~kpc, a molecular cloud of mass $10^4 M_{\odot}$ can emit TeV gamma rays at a detectable level if it is located within few hundred parsecs from the SNR \citep{gabici2}. This implies that the angular displacement between the SNR shell and the gamma ray emission is of the order of $\vartheta \approx 6^{\circ} (D/1 {\rm Kpc})^{-1} (d_{cl}/100 {\rm pc})$, where $d_{cl}$ is the distance between the SNR and the cloud. This translates in the fact that sometimes the association between SNRs and molecular clouds can be not so obvious, given  that the separation between the two objects can be even larger than the detector's field of view. Following this rationale, Gabici and Aharonian (2007) proposed that some of the unidentified TeV sources detected by HEGRA \citep{hegraunid}, H.E.S.S. \citep{hessunid} and MILAGRO \citep{milagrounid} might in fact be such clouds illuminated by a nearby SNR.

In this paper, we calculate the expected non-thermal emission, from radio to multi-TeV photons, from a molecular cloud illuminated by CRs coming from a nearby SNR. We generalize the model presented by Gabici and Aharonian (2007), which was limited to the hadronic TeV photons only, to include the generation of secondary electrons in the cloud and the related synchrotron and Bremsstrahlung emission. 
We found that the total radiation energy output from a cloud is dominated by the gamma ray emission, which can exceed the emission in other bands by an order of magnitude or more. This suggests that some of the unidentified TeV sources detected so far, with no obvious or very weak counterparts in other wavelengths (the so called "dark sources"), might be in fact associated with massive molecular clouds illuminated by CRs. Moreover, under certain conditions, the gamma-ray spectrum from the cloud exhibit a concave shape, being steep at low ($\sim$ GeV) energies and hard at high ($\sim$ TeV) energies. This fact might have important implications for the studies of the spectral compatibility of GeV and TeV gamma ray sources. 

\section{The model} \label{sec:model}

Consider a supernova of total energy $10^{51} E_{51} \, {\rm erg}$ exploding in a medium of density $n$.  
The initial shock velocity is $10^9 u_9$ cm/s
and remains roughly constant until the mass of the swept up material equals the mass of the ejecta.
This happens at a time  $t_{Sedov} \approx 200 [E_{51}/(n \, u_9^5)]^{1/3} {\rm yr}$, when the shock radius is $\approx 2.1 [E_{51}/(n \, u_9^2)]^{1/3} {\rm pc}$.
Then the SNR enters the Sedov phase 
and the shock radius and velocity scales with time as 
$R_{sh} \propto t^{2/5}$ and $u_{sh} \propto t^{-3/5}$.
%\begin{equation}
%R_{sh}(t) = R_{ad} \left( \frac{t}{t_{ad}} \right)^{\frac{2}{5}} 
%~~,~~
%u_{sh}(t) = \frac{d R_{sh}(t)}{dt} \propto t^{-\frac{3}{5}}
%\end{equation}

The spectrum of particles accelerated at the SNR shock is determined by the transport equation \citep[e.g.,][]{drury}:
\begin{equation}
\label{eq:transport}
\frac{\partial f}{\partial t} - \nabla D \nabla f + {\bf u} \nabla f - \frac{\nabla {\bf u}}{3} p \frac{\partial f}{\partial p} = 0 .
\end{equation}
where $D = D(p)$ is the momentum dependent diffusion coefficient and ${\bf u}$ the flow velocity.
For a strong shock with compression factor $r_s = 4$, the test particle theory predicts an universal shape for the CR spectrum at the shock $f_0(p) \propto p^{-4}$ \citep{drury}. If the shock is an efficient accelerator (as SNR shocks are believed to be), the CR pressure modifies the flow structure, making the shock more compressible and the spectrum of the accelerated particles harder, $f_0(p) \propto p^{-\alpha}$ with $3.5 \la \alpha \leq 4$ \citep{malkovdrury}.  
Detailed calculations compared with multiwavelength observations of SNRs suggest the values $r_s \approx 7$ and $\alpha \approx 3.7$ \citep{don,berezhko}, which we adopt in the following.

The maximum momentum of the accelerated particles is determined by a confinement condition, namely that 
the diffusion length $l_d$ of the particles cannot exceed the characteristic size of the system $R_{sh}$: %. Following \citet{ptuskin}, we can write:
%\citep{ptuskin}:
\begin{equation}
\label{eq:pmax}
l_d = \frac{D(p_{max})}{u_{sh}} \la R_{sh} .
\end{equation}
The maximum possible energies are achieved when the acceleration proceeds in the Bohm diffusion limit, $D \propto p/B_{sh}$, with $B_{sh}$ the magnetic field strength at the shock. In this case the maximum momentum decreases with time as $p_{max}(t) \propto B_{sh} t^{-1/5}$. In fact, the drop of $p_{max}$ is even faster, given that the magnetic field is also expected to decrease with time.
This implies that at any time, particles with momentum above $p_{max}(t)$ quickly escape the remnant, generating a cutoff in the spectrum. The spectrum of the runaway particles can be approximated as a $\delta$--function \citep[see][]{ptuskin}:
$$
q_{esc}(p,t) = -\delta(p-p_{max})
$$
\begin{equation}
\times \int_0^{\infty} d^3 R \left( \frac{\partial p_{max}}{\partial t} 
+ \frac{\nabla {\bf u}}{3} p_{max} \right)
f(p_{max},R)
\end{equation}
where the integration has to be performed where the integrand is negative.
Thus, to calculate the flux of the runaway particles one has to know: \textit{(i)} the CR particle distribution function at $p_{max}$ at any location in the SNR,
\textit{(ii)} the flow velocity both inside the shock and outside it, where the CR precursor forms and \textit{(iii)} how the maximum momentum varies during the SNR evolution.
Ptuskin and Zirakashvili (2005) showed that it is straigthforward to derive \textit{(i)} and \textit{(ii)} using an approximate (but still reasonably accurate) linear velocity profile inside the SNR :
\begin{equation}
u = \left( 1-\frac{1}{r_s} \right) \frac{u_{sh}(t)}{R_{sh}(t)} R
\end{equation}
and assuming that the CR pressure at the shock $P^{CR}_{sh}$ is a fraction $\xi_{CR}$ of the incoming ram pressure $\varrho u_{sh}^2$ and that $f_0(p_{max}) \propto P^{CR}_{sh}$.

The determination of $p_{max}$ and its evolution with time requires the knowledge of the diffusion coefficient (see Eq. \ref{eq:pmax}), which is in turn determined by the level of magnetic turbulence generated by the accelerated particles themselves. This makes the problem nonlinear and very difficult to be solved. The value of $p_{max}$ depends on a few crucial but poorly known aspects of the problem, including the nature of CR--driven instability operating in the shock precursor and the level of wave damping \citep{bell,bell2,ptuskin,pasquale,vladimirov}. 
Because of these uncertainties, we adopt here a phenomenological approach, namely we parametrize the maximum momentum as $p_{max}(t) \propto t^{-\delta}$. We further assume $p_{max} \sim 5$ PeV and $\sim 1$ GeV at the early ($t = 200$ yr) and late ($t = 5 \times 10^4$ yr) epochs of the Sedov phase respectively. This requires $\delta \approx 2.48$. 
Remarkably, if the maximum momentum is a power law function of time, the spectrum of the escaping particles integrated over the whole Sedov phase is also a power law of the form $\propto p^{-4}$ \citep{ptuskin}, which is close (sligthly harder) to what needed to fit the CR data below the knee \citep{CRbook}.

Following Ptuskin and Zirakashvili (2003), the approximate spectrum of the CRs inside the SNR $f_{in}(R,p,t)$ can be obtained from Eq. \ref{eq:transport} by dropping the diffusion term, while the spectrum of the runaway CRs at a given distance $R$ from the SNR and at a given time $t$ can be obtained by solving the diffusion equation:
\begin{equation}
\label{eq:diffusionISM}
\frac{\partial f_{out}}{\partial t}(R,p,t) = D_{ISM}(p) \nabla^2 f_{out}(R,p,t) + q_{esc}(p,t) \delta(R) .
\end{equation}
The diffusion coefficient $D_{ISM}(p)$ describes the propagation of CRs in the galactic disk. The available CR data require a power--law energy dependence, $D_{ISM}(E) \propto E^{-s}$, with $D_{ISM} \approx 10^{28} {\rm cm}^2/{\rm s}$ at $E \approx 10$ GeV and $s \approx 0.3 \div 0.7$ \citep{CRbook}.
The constraints on the diffusion coefficient are obtained from the comparison between diffusion models and CR data and have to be considered as average galactic values. However, the conditions might be rather different in regions close to CR sources, in particular due to the presence of strong gradients in the CR distribution, which may enhance the generation of plasma waves and thus suppress the diffusion coefficient \citep{wentzel,vladvlad}. %Below we assume $D_{ISM} = 3 \times 10^{29} (E/1 PeV)^{0.5} {\rm cm}^2/{\rm s} $.  
The change in $s$ within the allowed range or the choice of a different normalization for $D_{ISM}$ does not alter qualitatively the results, the main effect being that the characteristic time scales of the problem change proportional to $1/D_{ISM}$.

Remarkably, if $p_{max}(t)$ scales as a power law of time, Eq. \ref{eq:diffusionISM} can be solved analytically and the distribution function of escaping cosmic rays at any given distance $R$ from the SNR and at any given time $t$ reads, for energies $E \ge c \times p_{max}(t)$:
\begin{equation}
\label{eq:CRout}
f_{out}(t,R,E) = \frac{\eta E_{SN}}{\pi^{3/2} \ln(E_{MAX}/E_{MIN})}  \; \frac{e^{-(\frac{R}{R_d})^2}}{R_d^3} \; E^{-2}
\end{equation}
where $E_{SN}$ is the total supernova explosion energy, $\eta$ is the fraction of such energy converted into CRs and $E_{MAX}$ and $E_{MIN}$ are the maximum and minimum energies of CRs accelerated during the Sedov phase.
The diffusion distance for a CR of energy $E$ is:
\begin{equation}
R_d(E) = \sqrt{4 D(E) (t-\chi(E))}
\end{equation}
where
\begin{equation}
\chi(E) = t_{Sedov}\left(\frac{E}{E_{MAX}}\right)^{-1/\delta}
\end{equation}
represents the time after the supernova explosion at which CRs with energy $E$ are released in the interstellar medium.
The solution derived by Ptuskin and Zirakashvili (2005) for the total CR spectrum %($\propto E^{-2}$)
injected by a SNR in the interstellar medium during the whole Sedov phase can be easily derived by integrating Eq.~\ref{eq:CRout} over space.
Finally, it has to be noticed that the total CR spectrum at a given time and at a given distance from the SNR is the sum of two contributions: {\it i)} a time dependent contribution from CRs coming from the SNR, whose spectrum is described by Eq.~\ref{eq:CRout} and {\it ii)} a steady contribution from the galactic CR background.    

Following the procedure described above, it is possible to evaluate, for any given time, the CR spectrum in proximity of a molecular cloud located at a given distance from the SNR. 
If the diffusion coefficient inside the cloud is not significantly smaller than the Galactic one, CR freely penetrate the cloud and the CR spectrum inside the cloud is not affected by propagation effects. Conversely, if the diffusion coefficient is significantly (more than one order of magnitude) reduced, low energy CRs are excluded from the cloud and a low energy cutoff appears in the CR spectrum, at an energy that depends on the value of the diffusion coefficient \citep[see][for details]{gabici1}.
Here, we assume free penetration of CRs and we refer the reader to Gabici et al. (2007) (and references therein) for a detailed discussion on CR exclusions from molecular clouds.
We do not consider here any contribution from CR electrons coming from the SNR, since they do not escape the remnant due to diffusive confinement (for low energy electrons) and severe synchrotron losses in the strong magnetic field (for high energy electrons). 

CR protons propagating inside a molecular cloud produce secondary electron-positron pairs during inelastic interactions in the dense intercloud medium. We calculate the spectrum of the injected secondary electrons $Q_e(t,E)$ by using the parameterization given by  Kelner et al. (2006) and we follow the time evolution of the electron distribution function $f_e(t,E)$ by solving the equation:
\begin{equation}
\frac{\partial f_e(t,E)}{\partial t} = \frac{\partial}{\partial E} \left[ \left(\frac{dE}{dt}\right)_e f_e(t,E)\right] + Q_e(t,E) - \frac{f_e(t,E)}{\tau_{esc}^e}
\end{equation}
where $(dE/dt)_e = E/\tau_{loss}^e$ is the energy loss rate for electrons, $\tau_{loss}^e$ the energy loss time, and $\tau_{esc}^e$ the diffusive escape time from the cloud.
All these time scales will be discussed and estimated in section 4.
Once the proton and electron CR spectra have been derived, the non-thermal radiation from the molecular cloud can be calculated. 

%The function %s $f_{in}$ and 
%$f_{out}$ can be used to evaluate the production rates of $\gamma$-rays, electron-positron pairs and neutrino due to CR interactions in the ambient gas surrounding the SNR. %both from the SNR itself and surrounding dense environments (e.g. from nearby molecular clouds).

\section{Cosmic ray spectrum at the cloud location}

\begin{figure*}
\resizebox{\hsize}{!}{
\includegraphics{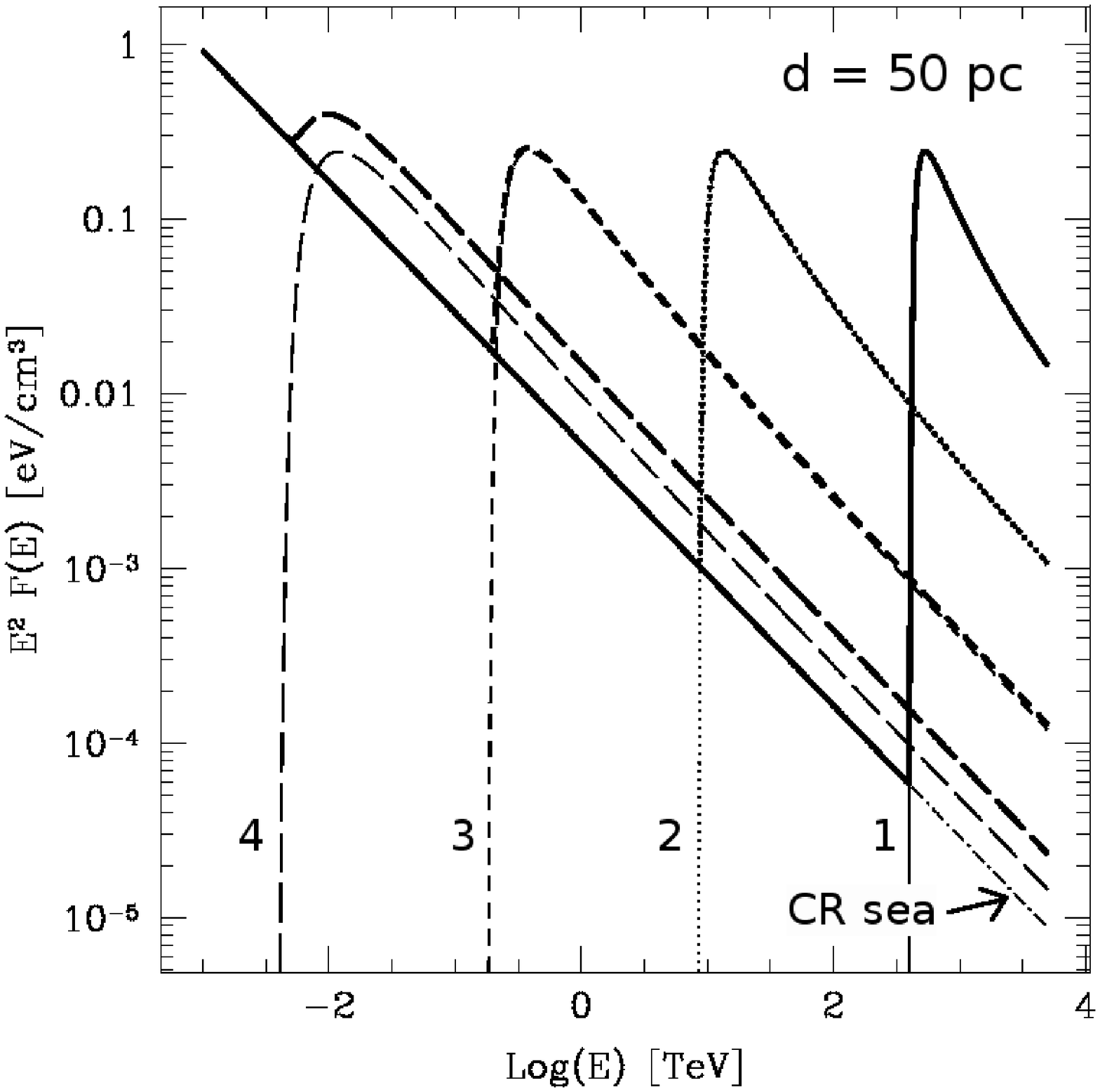}
\includegraphics{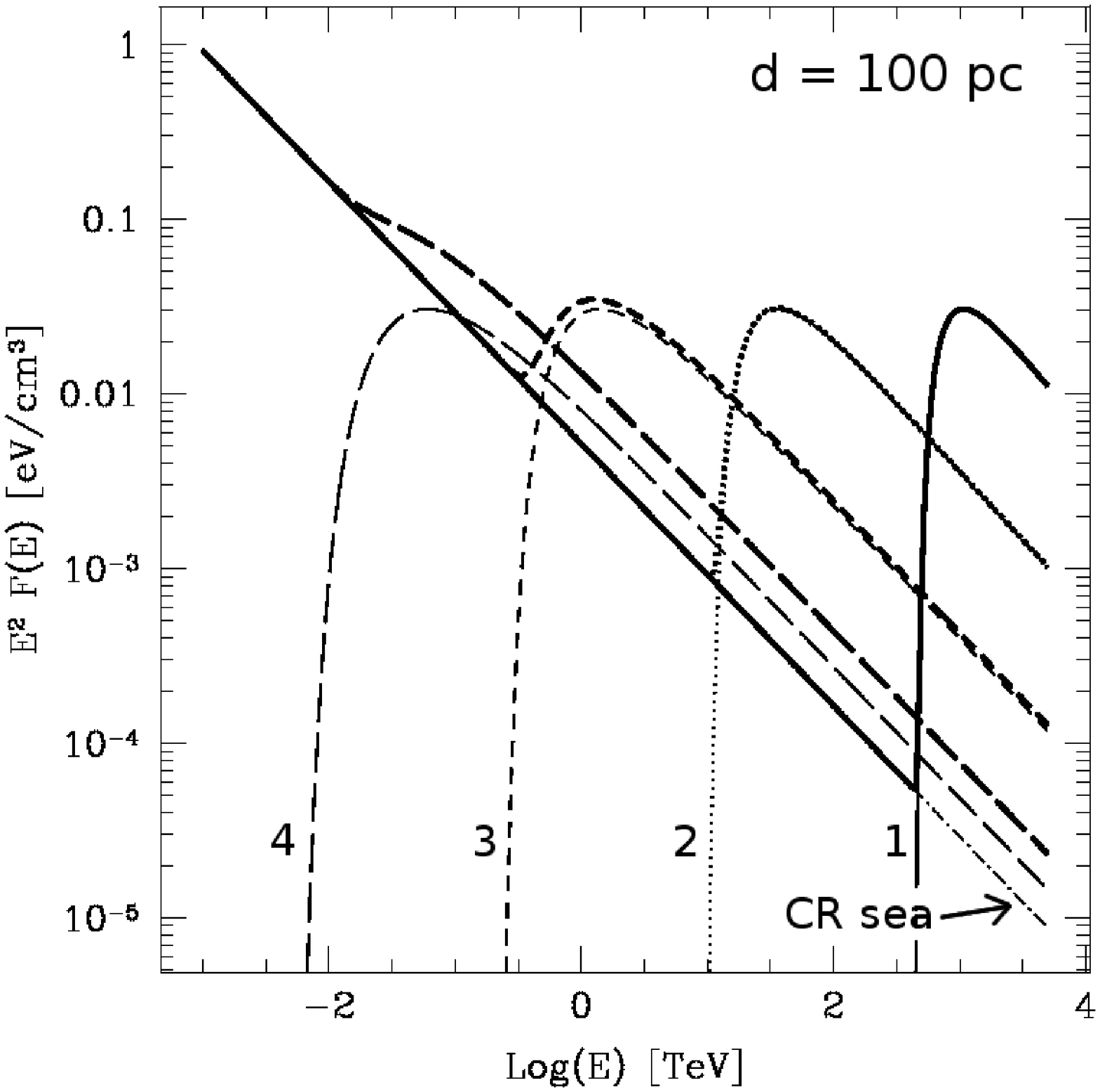}
} 
\caption{Spectrum of CRs at the location of the molecular cloud. The cloud is located at 50 pc (left) and 100 pc (right panel) from a SNR. The thin dot--dashed line shows the Galactic CR spectrum, while the thin solid (curve 1), dotted (2), short--dashed (3) and long-dashed (4) lines represent the spectrum of CRs coming from the SNR for 500, 2000, 8000 and 32000 years after the supernova explosion, respectively. The thick lines show the total CR spectrum at the cloud location.}
\label{fig:crs}
\end{figure*}

We consider here a molecular cloud located at a given distance $d_{cl}$ from a SNR and we calculate the CR spectrum at the cloud location. 
As mentioned  above, the spectrum is the sum of two disctinct components: {\it i)} the CRs coming from the nearby SNR, described by Eq.~\ref{eq:CRout}, and {\it ii)} the galactic CR background, which results from the superposition of all the CR sources in the Galaxy.
While the latter contribution is constant in time, the first one changes, since the flux of CRs escaping from the SNR evolves in time as described in Sec.~\ref{sec:model}.

Fig. \ref{fig:crs} shows the spectrum of CRs at the location of the  molecular cloud. The Galactic CR background is plotted as a thin dot--dashed line labeled as {\it CR sea}, while the spectrum of the CRs coming from the SNR is plotted as a thin line for different times after the supernova explosion: 500 yr (solid), 2000 yr (dotted), 8000 yr (short -- dashed) and 32000 yr (long--dashed).
Thick lines represent the sum of the two contributions.  
The distance between the SNR and the molecular cloud is 50 pc (left panel) and 100 pc (right panel).
We assume a total supernova explosion energy of $10^{51}$ ergs and a high CR acceleration efficiency at the SNR shock equal to $\eta = 30 \%$. 
The normalization of the CR spectrum at the cloud scales linearly with these two quantities \footnote{This is true only as a first approxiamation for the parameter $\eta$, since the estimated flux of particles escaping the SNR comes from a nonlinear theory of shock acceleration}.
The diffuse Galactic cosmic ray spectrum is assumed to be identical to the one observed near the Sun \citep[see e.g.][]{dermer}:
\begin{equation}
J_{CR}(E) = 2.2 \left( \frac{E}{GeV} \right)^{-2.75} {\rm cm^{-2} s^{-1} GeV^{-1} sr^{-1}}
\end{equation}
while the diffusion coefficient, needed to evaluate Eq.~\ref{eq:CRout}, is taken equal to:
\begin{equation}
\label{eq:diff}
D(E) = 10^{28} \left( \frac{E}{10 ~ GeV} \right)^{0.5} ~ cm^2/s ~ ,
\end{equation}
compatible with CR propagation models \citep[e.g.][]{CRbook}.

The evolution with time of the CR spectrum at the position of the molecular cloud can be understood as follows. 
According to the model described in Sec.~\ref{sec:model}, CRs with different energies leave the SNR at different times. 
The highest energy ($\sim PeV$) CRs leave the SNR first, while CRs with lower and lower energy are released at later times. 
Moreover, higher energy CRs diffuse faster, thus the spectrum of CRs at the cloud exhibit a sharp low energy cutoff at an energy $E_{low}$, which moves to lower and lower energies as time passes. 
The position of the cutoff represents the energy of the least energetic particles that had enough time to reach the cloud. 

From Fig.~\ref{fig:crs} it is clear that the influence of the presence of a nearby SNR close to the cloud is reflected in the CR spectrum at the cloud position, but this influence depends on many parameters, such as the distance between the SNR and the cloud, the time since the supernova explosion, and the CR energy.
The CRs coming from the SNR dominate the total CR spectrum at high energy, while at lower ($\sim$ GeV) energies the galactic CR background is always the dominant component, unless the molecular cloud is located at distances significantly smaller than $\approx$ 50 pc from the SNR.
However, such small distances are comparable to the size of the SNR itself and thus in this case an interaction between the SNR shock and the  molecular cloud is expected \citep[see e.g.][]{dav2}.
The investigation of this scenario goes beyond the scope of this paper and thus we limit ourselves to the case in which the distance between the SNR and the cloud is significantly bigger than the size of both objects.

%it is not possible to classify a molecular cloud as {\it active} or {\it passive} in a unambiguous way, the classification depending on the CR energy, on the time after the supernova explosion and on the distance between the SNR and the cloud.
%This issue will be discussed in detail in the next sections, when the non-thermal flux from a molecular cloud illuminated by CRs will be calculated.

Aharonian and Atoyan (1996) also evaluated the CR spectrum in the vicinity of a CR accelerator by using an approach similar to the one developed here.
They made no specific assumption about the nature of the accelerators and solved the CR transport equation by assuming that a power law spectrum of CRs is injected in the interstellar medium.
They considered both the case of a continuous injection of particles during the whole lifetime of the accelerator and the case of an impulsive source that releases all the CRs at the same time.
The approach we adopt here is different, because it is specific for a given class of sources, namely SNRs. In this specific case, particles having different energies are released at different times in the interstellar medium.

\section{Relevant time scales for cosmic ray propagation inside a molecular cloud}

Molecular clouds are characterized by a wide range of masses, going from $\approx 10 ~ M_{\odot}$ to $10^5 M_{\odot}$ or even more and have typical sizes ranging from few to few tens of parsecs. The typical density of a cloud is of about few hundred atoms per cubic centimeter, but much higher densities can be found in less massive and smaller (parsec or sub-parsec scale) molecular cloud cores, dark clouds or Bok globules \citep[see][for a review]{palla}.
The typical magnetic field of the intercloud medium is $\approx 10 ~ \mu$G \citep{shu}, and it scales roughly as the square root of the gas density, thus reaching the mG level in the densest regions with density $10^5 \div 10^6$~cm$^{-3}$ \citep{crutcher}.

The propagation of high energy CRs inside molecular clouds has been studied in Gabici et al (2007) (see also Dogel' and Sharov, 1990), where an extensive discussion can be found. Here we summarize the most relevant aspects.
Once the CRs from the SNR reach the molecular cloud, they diffusively penetrate with typical time scale:
\begin{equation}
\label{eq:difftime}
\tau_{diff} \sim \frac{R_{cl}^2}{6 D(E,B)}
\end{equation}
where $R_{cl}$ is the cloud radius and $D$ is the diffusion coefficient which we assume here to depend on energy and on magnetic field as:
\begin{equation}
\label{eq:diffcoeff}
D(E) = \chi  ~ 10^{28} \left( \frac{E}{10 ~ GeV} \right)^{0.5}  \left( \frac{B}{3 ~ \mu G} \right)^{-0.5}  ~ cm^2/s ~ .
\end{equation}
Here $\chi$ is a factor that takes into account deviations from the average Galactic diffusion coefficient described  by Eq.~\ref{eq:diff}  \citep[see][]{gabici1} and $3 ~ \mu$G is the average magnetic field in the Galactic disk.
At very high energies, when the Larmor radius of the particle becomes comparable with or even larger than the size of the cloud, particles propagate almost rectilinearly, and the propagation time reduces to the crossing time $\tau_{cross} = R_{cl}/c$.
In the following we assume the characteristic propagation time for a CR in a molecular cloud to be: 
\begin{equation}
\tau_{prop} \approx \tau_{diff}+\tau_{cross},
\end{equation}
which is a rough approximation which still describes with sufficient accuracy the two different regimes of propagation.

CRs can freely penetrate the molecular cloud if the diffusion time is shorter than the energy loss time which, for CR protons with energy above $\sim 1$GeV, is dominated by inelastic proton--proton interactions in the dense gas and reads \citep[see e.g.][]{CRbook}:
\begin{equation}
\label{eq:pptime}
\tau_{pp} = \frac{1}{n_{gas} c \kappa \sigma_{pp}} = 6 \times 10^5 \left(\frac{n_{gas}}{100~cm^{-3}} \right)^{-1} {\rm yr},
\end{equation}
where $n_{gas}$ is the gas density, $c$ is the speed of light and $\sigma_{pp}$ and $\kappa$ are the cross section and inelasticity of the process, respectively.
For nonrelativistic protons (energies below 1~GeV) ionization losses become relevant, with time scale \citep[e.g.][]{CRbook}:
$$
\tau_{ion}^p \sim 2.8 \times 10^7 \left(\frac{n_{gas}}{100~cm^{-3}}\right)^{-1} \times 
$$
\begin{equation}
\label{eq:iontimep}
\left(\frac{E_k}{m_pc^2}\right)^{\frac{3}{2}}\left[11.8+ln\left(\frac{E_k}{m_pc^2}\right)\right]^{-1} ~ yr .
\end{equation}
Here $E_k$ is the proton kinetic energy and $m_p$ is the proton mass.
The total energy loss time for CR protons due to both processes is given by:
\begin{equation}
\tau_{loss}^p = \left(\frac{1}{\tau_{pp}}+\frac{1}{\tau_{ion}^p}\right)^{-1}.
\end{equation}

\begin{figure*}  
\resizebox{\textwidth}{!}{
\includegraphics{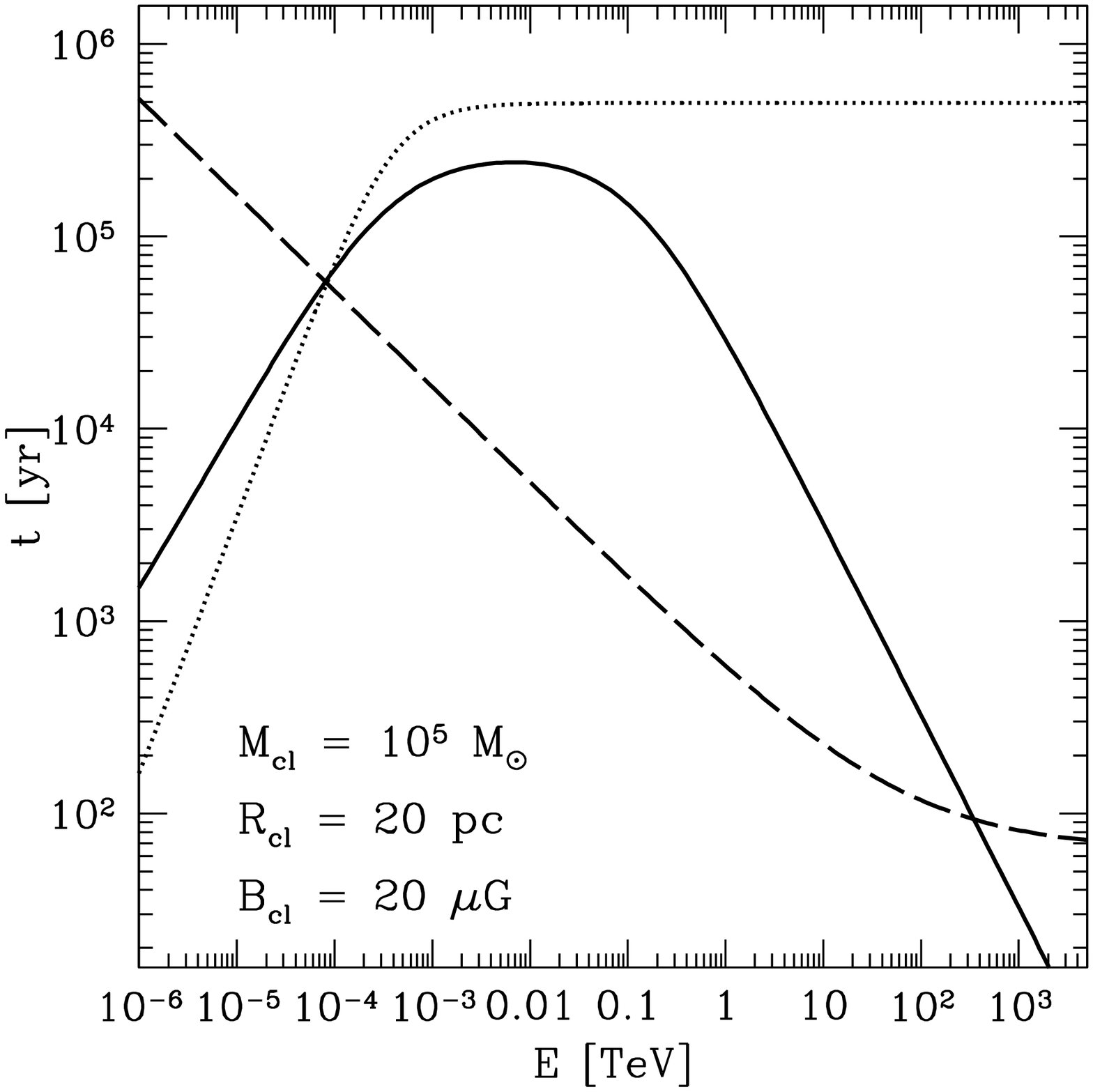}
\hspace{-1.5cm}
\includegraphics{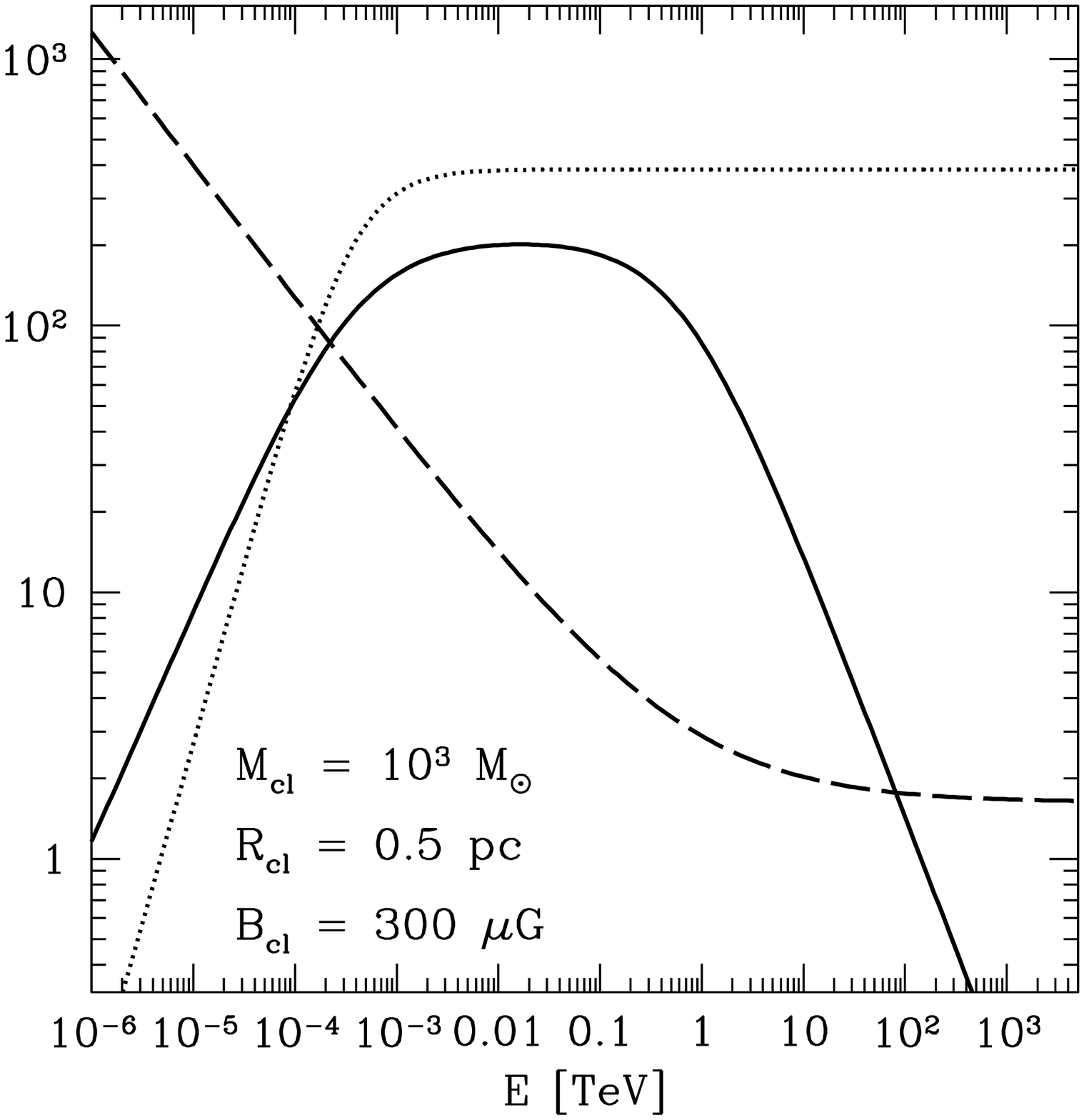}
\hspace{-1.5cm}
\includegraphics{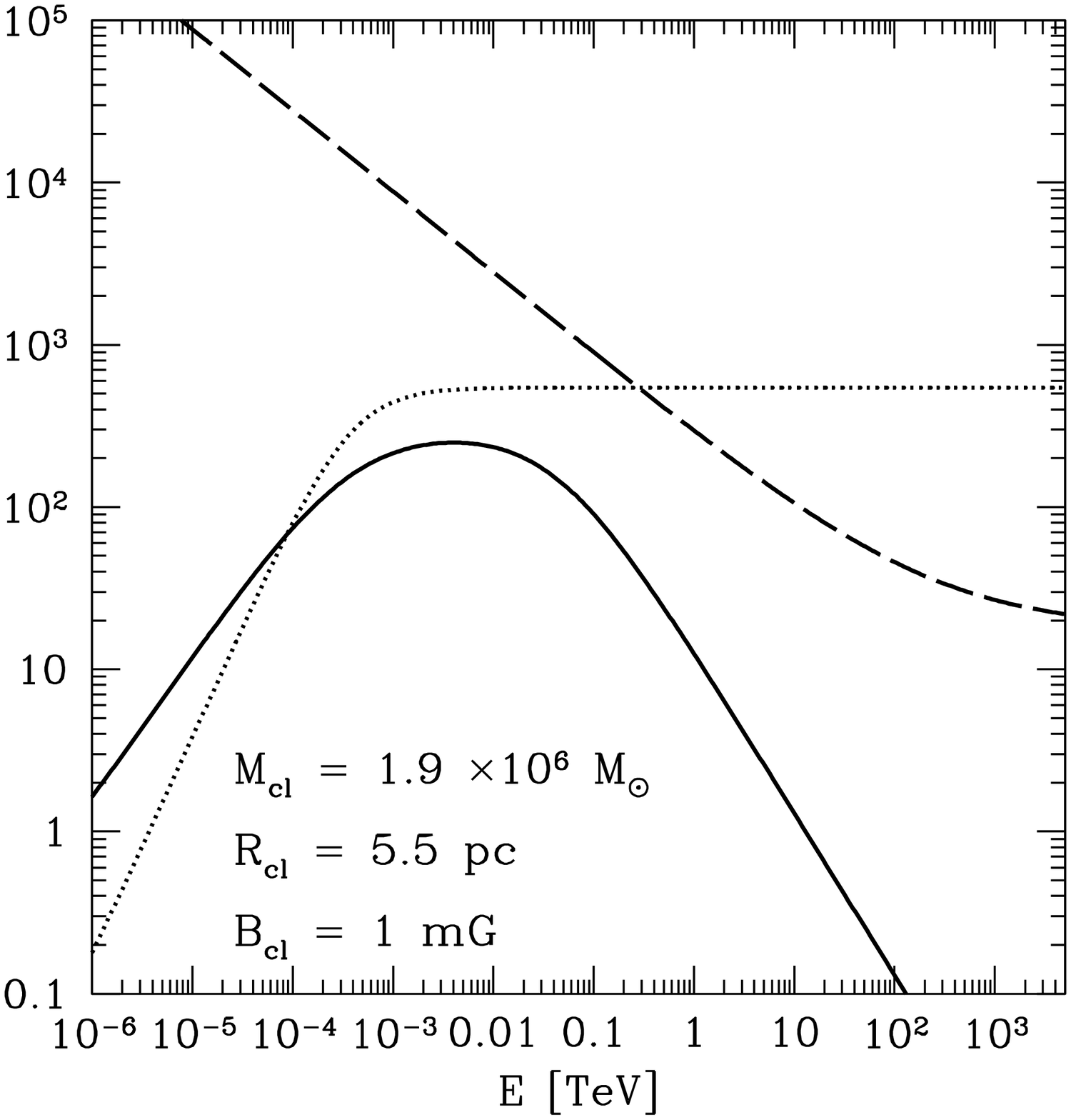}
} 
\caption{{\bf LEFT PANEL:} relevant time scales for CR propagation inside a molecular cloud with mass $M_{cl} = 10^5 M_{\odot}$, radius $R_{cl} = 20$~pc, and magnetic field $20~\mu$G. Assuming a flat density profile gives a density of $\sim 120$~cm$^{-3}$. The dashed line represents the CR propagation time scale over a distance $R_{cl}$. The dotted line represents the energy loss time for CR protons (ionization losses are relevant below 1 GeV, inelastic proton-proton interaction at higher energies) while the solid line refers to the energy loss time for CR electrons, including ionization losses, Bremsstrahlung losses and synchrotron losses, which dominates at low, intermediate and high energies respectively. {\bf MIDDLE PANEL:} same as left panel, but for a cloud with mass $M_{cl} = 10^3 M_{\odot}$, radius $R_{cl} = 0.5$ pc, and magnetic field $300~\mu$G. The corresponding density is $\sim 1.6 \times 10^5$cm$^{-3}$. {\bf RIGHT PANEL:} same as the left panel, but for the specific case of the SgrB2 cloud with mass $1.9 \times 10^6 M_{\odot}$, radius 5.5 pc and magnetic field 1 mG. This implies an average density of $1.1 \times 10^5$ cm$^{-3}$ (see text for an explanation of the choice of parameters adopted).}
\label{fig:timeGMC}
\end{figure*}

Secondary electron-positron pairs are produced inside the cloud during inelastic collisions between CR protons in the dense gas. The typical diffusion time for such electrons is given by Eq.~\ref{eq:difftime}, while the relevant channels for electron energy losses are ionization losses, Bremsstrahlung emission and synchrotron emission with characteristic time scales \citep[e.g.][]{ginzburg}:
\begin{equation}
\label{eq:iontimee}
\tau^e_{ion} = 1.9 \times 10^4 \left(\frac{n_{gas}}{100~cm^{-3}}\right)^{-1} \frac{\gamma}{3 ~ ln(\gamma)+18.8} ~ yr
\end{equation}
\begin{equation}
\label{eq:bremstime}
\tau_{Brems} = 3.3 \times 10^5 \left(\frac{n_{gas}}{100~cm^{-3}}\right)^{-1} ~ yr 
\end{equation}
\begin{equation}
\label{eq:syntime}
\tau_{syn} = 1.3 \times 10^5 \left(\frac{E}{TeV}\right)^{-1} \left(\frac{B}{10 ~ \mu G}\right)^{-2} ~ yr
\end{equation}
respectively. 
Here, $\gamma$ is the electron Lorenz factor and $E$ the total electron energy.
The ionization losses are computed for ultra-relativistic electrons. 
The total energy loss time for CR electrons due to the three processes listed above is given by:
\begin{equation}
\tau_{loss}^e = \left(\frac{1}{\tau_{ion}^e}+\frac{1}{\tau_{Brems}}+\frac{1}{\tau_{syn}}\right)^{-1}.
\end{equation}
In writing the equation above, we neglected the possible role of inverse Compton losses off soft background photons. 
Such losses are expected to be relevant when the energy density in the radiation field $\omega_{rad}$ is greater than the magnetic field energy density. In the following we will consider a cloud with radius $R_{cl} = 20$ pc and magnetic field $B = 20 \mu$G, and in this case the condition reads: $\omega_{rad} > 10 ~ (B/20 ~ \mu G)^2 ~  {\rm eV}/{\rm cm}^3$. Such condition can be realized inside molecular clouds only close to star forming regions. In particular, if a star forming region or a stellar OB association with total photon output equal to $4 \times 10^{33} L/L_{\odot}$~erg/s is located within a molecular cloud, inverse Compton losses will dominate over synchrotron losses around the star forming region up to a distance of  $R \approx 8 \times 10^{-3} (L/L_{\odot})^{1/2} (B/20 ~ \mu G)^{-1}$ pc, which becomes comparable with or larger than the molecular cloud radius $R_{cl}$ when $L \ge 6 \times 10^6 (B/20~\mu G)^2 (R_{cl}/20~{\rm pc})^2 L_{\odot}$. 
Since this is a quite high luminosity, only a small fraction of molecular clouds are expected to host such luminous OB associations (Williams and McKee 1997). We can thus safely neglect the role of inverse Compton losses in most of the situations.
Inverse Compton losses can certainly play a role if one considers small regions within the cloud which surrounds star forming regions. Such regions constitute a small fraction of the cloud volume and thus are not expected to affect significantly the non thermal emission from the whole cloud.

In Fig.~\ref{fig:timeGMC} (left panel) the characteristic time scales listed above have been plotted as a function of particle energy for a giant molecular cloud with total mass $M_{cl} = 10^5 M_{\odot}$ and radius $R_{cl} = 20$ pc. Assuming a flat density profile the density is $n_{gas} \sim 120$ cm$^{-3}$. The magnetic field is assumed to be $B_{cl} = 20~ \mu$G. The dotted line refers to proton energy losses, which are dominated by ionization losses at energies below $\sim 1$ GeV and by inelastic proton--proton interactions at higher energies. The solid line represents the electron energy loss time. The three different power law behaviors reflect the dominance of ionization, Bremsstrahlung and synchrotron losses at low, intermediate and high energies, respectively. Finally, the dashed line represents the propagation time over a distance equal to the cloud radius. The propagation time has been evaluated by using the diffusion coefficient from Eq.~\ref{eq:diffcoeff} with $\chi = 1$ (no suppression with respect to the average Galactic value).

For proton energies above the threshold for pion production ($E_{th} \sim 280$ MeV), the propagation time is always shorter than the energy loss time. This means that CR protons which produce gamma rays and secondary electrons can freely penetrate the cloud and their flux is not attenuated due to energy losses.
The propagation time for CR electrons is also shorter than the energy loss time for particle energies between $E \sim 100$ MeV and a few hundreds of TeV. This implies that, within this energy range, the secondary electrons produced inside the cloud quickly escape, and have little effect on the non-thermal emission from the cloud. On the other hand, extremely energetic electrons with energies above a few hundreds TeVs radiate all their energy in form of synchrotron photons before leaving the cloud. In a typical magnetic field of a few tens of microGauss, these electrons emit synchrotron photons with energy:
\begin{equation}
E_{syn} \approx 1 \left(\frac{B_{cl}}{10~\mu {\rm G}}\right) \left(\frac{E}{100 ~ {\rm TeV}}\right)^2 ~ {\rm keV} ~ .
\end{equation}
Thus, the most relevant contribution from secondary electrons to the cloud non thermal emission falls in the hard X-ray band. 

The middle panel of Fig.~\ref{fig:timeGMC} shows the time scales for a compact cloud with radius $R_{cl} = 0.5$~pc and mass $M_{cl} = 10^3 M_{\odot}$, which implies, in case of a flat density profile, a density of $\sim 1.6 \times 10^5$~cm$^{-3}$. A strong magnetic field of $300~\mu$G is assumed, as suggested by observations of very dense clouds \citep{crutcher}. These parameters are typical for molecular cloud cores or for dark clouds \cite{palla}. The left and middle panels in Fig.~\ref{fig:timeGMC} looks very similar, except for a scaling factor. This implies that the same conclusions can be drawn as for the case of a giant molecular cloud, and thus a similar behavior is expected from giant molecular clouds and compact clouds, the only difference being that all the time scales are much shorter in the latter case, due to the high gas density and to the reduced size of the system.

The properties of giant molecular clouds located in the galactic centre region can differ significantly from the average figures reported above.
As an example, we plot in the right panel of Fig.~\ref{fig:timeGMC} the typical time scales for the SgrB2 cloud.
This is a very massive cloud located at 100 pc (projected distance) from the galactic centre. 
The cloud virial mass is $M_{SgrB2} = 1.9 \times 10^6 M_{\odot}$ \citep{protheroe} and a magnetic field at the milliGauss level has been measured in the outer envelope of the cloud complex \citep{crutcher}.
The mass distribution can be fitted with a radial gaussian density profile with $\sigma = 2.75$ pc \citep[][and references therein]{protheroe}.
To compute the curves plotted in Fig.~\ref{fig:timeGMC} (right panel), we assumed a cloud radius of $R_{SgrB2} = 2 \sigma = 5.5$~pc, which encloses $\approx 95\%$ of the total cloud mass. This gives an average density of $n_{gas} = 1.1 \times 10^5$cm$^{-3}$ (roughly a factor of 2 below the central density).
It is evident from the right panel of Fig.~\ref{fig:timeGMC} that the SgrB2 cloud is remarkably different from a typical giant molecular cloud.
In particular, the very high values of the magnetic field and of the gas density make the energy loss time of CR protons significantly shorter than the propagation time for energies below a few hundred GeVs.
Moreover, for CR electrons the energy loss time is always shorter than the propagation time.
This means that CR protons with energies up to few hundred GeVs cannot penetrate the molecular cloud, as they do in the cases considered in the left and middle panel of Fig.~\ref{fig:timeGMC}.
Primary CR electrons cannot penetrate the cloud, while secondary CR electrons produced inside the cloud in hadronic interactions cannot leave the cloud and radiate all their energy close to their production site.
These characteristics make SgrB2 a very peculiar objects whose modelling needs a specific treatment.
A detailed study of the CR penetration in the SgrB2 cloud has been performed by Protheroe et al. (2008), who also computed the synchrotron radio emission from secondary electrons.
The effects of CRs exclusion from giant molecular clouds on their gamma ray emission have been discussed in detail by Gabici et al. (2007).
In the following we will not focus onto any specific object but rather investigate the case of the typical molecular clouds as the ones considered in the left and middle panels of Fig.~\ref{fig:timeGMC}.

To conclude the discussion on characteristic time scales, it is interesting to note that under certain conditions all the energy loss times, for both protons and electrons, scales in the same way with gas density, namely as $n_{gas}^{-1}$. This is evident from Eqns.~\ref{eq:pptime} and \ref{eq:iontimep}, that describe inelastic proton-proton scattering and ionization losses for protons respectively, and from Eqns.~\ref{eq:iontimee} and \ref{eq:bremstime} that describe ionization and Bremsstrahlung losses for electrons respectively. Moreover, also for synchrotron losses (Eq.~\ref{eq:syntime}) it is possible to derive the same scaling with gas density by recalling that observations suggest that the magnetic field in molecular clouds scales as the square root of gas density \citep{crutcher}.
This has the important consequence that, in typical molecular clouds, particles with a given energy (protons or electrons) lose their energy always through the same channel, independently on the gas density.  
As said above, this conclusion may not hold for peculiar obkects such as SgrB2.

\section{Non-thermal radiation from a molecular cloud}

Molecular clouds are now estabilished gamma ray sources \citep{galridge,w28}. 
Their gamma ray emission is believed to be the result of the decay of neutral pions produced during the inelastic collisions of CRs with the dense gas which constitutes the cloud \citep{issa,felixclouds,gabici1}.
During the same interactions, also electrons and positrons are produced via the decay of charged pions. These electrons and positrons produce a broad spectrum of radiation from radio waves to gamma rays due to synchrotron emission and non thermal Bremsstrahlung.

%If a molecular cloud is embedded in the ``sea'' of Galactic CRs, assumed to be constant throughout the Galaxy, and if CRs can freely penetrate into the cloud, then the total gamma ray flux above a given energy $E_{\gamma}$ is solely determined by the mass of the cloud and its distance to the Earth and can be written as \citep{felixclouds}:
%\begin{equation}
%F(\ge E_{\gamma}) \sim 1.45 \times 10^{-13} E_{TeV}^{-1.75} \left( \frac{M_5}{D_{kpc}^2} \right) {\rm cm^{-2} s^{-1}} 
%\end{equation} 
%where $M_5$ is the cloud mass in units of $10^5 M_{\odot}$ and $D_{kpc}$ is the cloud distance in kpc.
%This flux can be enhanced if the cloud is located in the vicinity of a CR accelerator, since in that case the cosmic ray spectrum at the cloud location might be enhanced with respect to the average Galactic one, due to the contribution from freshly accelerated CRs that escape the accelerator \citep{atoyan,gabici2}.

%Following the nomenclature adopted in the existing literature, we classify a cloud as {\it passive} if the CR spectrum at the cloud location is equal to the average Galactic one, and {\it active} if the CR level is enhanced with respect to the averace Galactic one due to the presence of a nearby CR accelerator. 

\begin{figure*}  
\resizebox{.80\textwidth}{!}{
\includegraphics{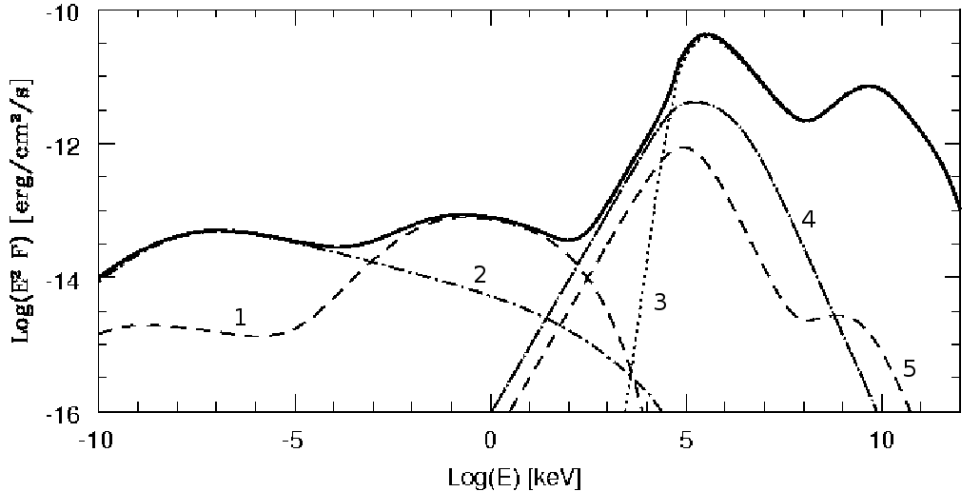}
} 
\caption{Broad band spectrum for a molecular cloud of mass $10^5 M_{\odot}$, radius $20$~pc, density $\sim 120$~cm$^{-3}$, magnetic field $20 \mu$G. The molecular cloud is at 100 pc from a SNR that exploded 2000 yr ago. The distance  of the cloud is 1 kpc. The dotted line shows the emission from $\pi^0$--decay (curve 3), the dot-dashed lines represent the synchrotron (curve 2) and Bremsstrahlung (curve 4) emission from background CR electrons that penetrate the molecular cloud and the dashed lines the synchrotron (curve 1) and Bremsstrahlung (curve 5) emission from secondary electrons.}
\label{fig:GMC2000}
\end{figure*}

In this section we compute the expected non-thermal emission from a molecular cloud located in the proximity of a SNR. The emission is the result of CR interactions with the dense gas and magnetic field in the cloud and is made up of two contributions: a steady state contribution from the interactions of background CRs that penetrate the cloud and a time dependent contribution from the interactions of CRs coming from the nearby SNR.

We consider a giant molecular cloud of mass $M_{cl} = 10^5 M_{\odot}$, radius $R_{cl} = 20$~pc and we assume an uniform density of $n_{gas} \sim 120$~cm$^{-3}$. The magnetic field is $B_{cl} = 20 ~ \mu$G. The relevant time scales for CR propagation and energy losses in such an environment have been plotted in the left panel of Fig.~\ref{fig:timeGMC}.
In order to show all the different contributions to the total non-thermal emission, in Fig.~\ref{fig:GMC2000} we plot the broad band spectrum from the cloud at a time $t = 2000$ years after the supernova explosion.
The SNR is located 100 pc away from the molecular cloud and the distance to the observer is 1 kpc.
%The dotted line represents the emission from neutral pion decay (both from background CRs and CRs from the SNR), the dot--dashed lines represent the synchrotron (leftmost curve) and Bremsstrahlung (rightmost curve) from background CR electrons that penetrate the cloud and the dashed line represents the synchrotron and Bremsstrahlung emission from secondary electrons produced during inelastic CR interactions in the dense gas.    
The dotted line (curve 3) represents the emission from neutral pion decay (from both background CRs and CRs from the SNR), the dot--dashed lines represent the synchrotron (curve 2) and Bremsstrahlung (curve 4) emission from background CR electrons that penetrate the molecular cloud and the dashed lines represent the synchrotron (curve 1) and Bremsstrahlung (curve 5) emission from secondary electrons produced during inelastic CR interactions in the dense gas.
For the spectrum of background CR electrons we use a fit to the measured spectrum at Earth (see Kobayashi et al. 2004 for a recent compilation of data). Electrons can freely penetrate the cloud except for the highest ($\ga 300 ~ {\rm TeV}$) and lowest ($\la 100 ~ {\rm MeV}$) part of the energy spectrum, where the energy loss time scale is significantly shorter than the propagation time (see Fig.~\ref{fig:timeGMC}).
For this energies the CR electron flux inside the cloud is suppressed and we estimated the suppression by assuming that CR electrons can penetrate undisturbed the cloud only up to a given depth, which is estimated as $\approx c \tau_{loss}^e$ and $\approx \sqrt{D \tau_{loss}^e}$ for the high and low energy end of the spectrum respectively. This approximation is satisfactory for the purposes of this paper, given that the suppression becomes crucial only for particles which emit negligible non-thermal emission. 

The decay of neutral pions dominates the total emission for energies above $\approx 100$~MeV. The two peaks in the emission reflects the shape of the underlying CR spectrum, which, as illustrated in Fig.~\ref{fig:crs}, is the sum of the steep background CR spectrum, which produces the $\pi^0$--bump at a photon energy of $m_{\pi^0}/2 \sim 70$~MeV (in the photon flux $F$), and an hard CR component coming from the SNR that produces the bump at higher energies. The flux level at 1 TeV is approximatively $5 \times 10^{-12}$erg/cm$^2$/s, detectable by currently operating Imaging Atmospheric Cherenkov Telescopes, even taking into account the quite extended ($\approx 2^{\circ}$) nature of the source. It is remarkable that such a cloud would be detectable even if it were located at the distance of the Galactic centre, as can be easily estimated by taking into account that the sensitivity of a Cherenkov telescope like H.E.S.S. after 50 hours of exposure, is $\approx 10^{-13} (\theta_s/0.1^{\circ})$TeV/cm$^2$/s, where $\theta_s$ is the source extension. This means that very massive clouds can be used to reveal the presence of enhancements of the CR density in different locations throughout the whole Galaxy. Similar conclusions can be drawn for the expected emission in the GeV range, which is currently probed by the AGILE and FERMI satellites. In particular, FERMI, with a point source sensitivity of $\la 10^{-9}$GeV/cm$^{2}$/s at energies above 1 GeV (www-glast.slac.stanford.edu), will be able to detect such giant molecular clouds as extended sources if they are located within $1 \div 2$ kpc from the Earth, or as point sources if they are at larger distances.
Such a use of molecular cloud as CR {\it barometers} has been discussed in several papers for both GeV \citep{issa} and TeV gamma rays \citep{felixclouds}.  Here we demonstrated that SNRs can provide enhancements in the CR density that can generate gamma ray fluxes well within the capabilities of currently operating instruments.

The spectral shape in the gamma ray energy range deserves further discussion. For the situation considered in Fig.~\ref{fig:GMC2000}, the GeV gamma rays are the result of the decay of neutral pions produced by background CRs that penetrate the cloud. Thus, the gamma ray spectrum above GeV energies simply mimics the underlying CR spectrum, which is a steep power law  of the form $\approx E^{-2.75}$.
On the other hand, the neutral pion decay spectrum at TeV energies is, in this case, dominated by the contribution from CRs coming from the nearby SNR. After 2000 years from the supernova explosion, only CRs with energies above several tens of TeV had enough time to leave the SNR and reach the cloud and thus the CR spectrum at the cloud exhibits an abrupt low energy cutoff at that energy, that we call here $E^{cut}_{CR}$ (see Fig.~\ref{fig:crs}). As a consequence, the gamma ray spectrum is expected to be peaked at an energy $\sim E_{CR}^{cut}/10$ of several TeV. The slope of the gamma ray spectrum below the peak is determined by the physics of the interaction only, and not by the shape of the underlying CR spectrum, and is roughly of the form $E^2 dN/dE \propto E^2$ \citep[e.g.][]{kelner}. 
Thus, a loose association between a SNR and a massive molecular clouds as the one studied here, is expected to be characterized, at least at some stage of the SNR evolution, by a very peculiar spectrum which is steep at low (GeV) energies and hard at high (TeV) energies.

\begin{figure*}  
\resizebox{.70\textwidth}{!}{
\includegraphics{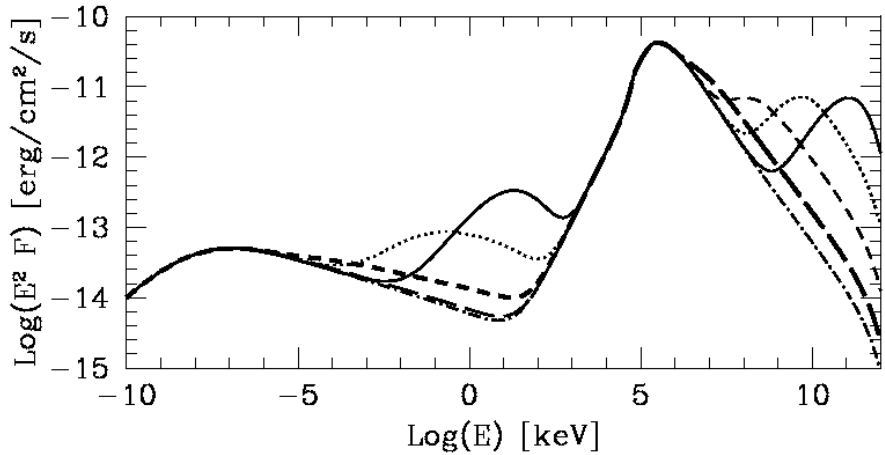}
} 
\caption{Time dependence of the broad band spectrum for the same giant molecular cloud considered in Fig.~\ref{fig:GMC2000}. The solid, dotted, short--dashed and long--dashed lines refers to the emission at a time 500, 2000, 8000 and 32000 years after the explosion. The dot--dashed line represent the spectrum of a molecular cloud with no SNR in its proximity.}
\label{fig:GMCtot}
\end{figure*}

The possibility of detecting sources with such a distinct spectrum is also relevant for the issue of identifying GeV and TeV unidentified sources.  
One of the criteria suggested to support an association between a GeV and TeV source is, beside the positional coincidence, the spectral compatibility.
In a recent paper,  Funk et al. (2008), investigated the spectral compatibility of EGRET and H.E.S.S. unidentified sources located in the inner Galactic region. For sources showing positional coincidence, they found generally a good spectral compatibility, but due to the small number of sources, it was not possible to claim an association between the two populations of sources at a statistically significant level. For sources detected by only one of the two instruments, Funk et al. (2008) investigated the consequences of extrapolating the measured spectrumat higher or lower energies. They considered only single power laws extending over the whole (GeV--TeV) energy range and discussed the implications of adding a high or low energy cutoff. 
The scenario presented here adds the new interesting possibility of a single source showing a dramatically different behavior at GeV and TeV energies, namely a spectrum which shows a significant hardening at higher energies.

The evolution with time of the emission from the cloud is shown in Fig.~\ref{fig:GMCtot}, where the solid, dotted, short--dashed and long--dashed lines show the spectrum at a time equal to 500, 2000, 8000 and 32000 years after the supernova explosion respectively. 
For comparison, the emission from a molecular cloud, with no SNR located in its proximity, is plotted as a dot--dashed line.
In this case, only background CRs that penetrate the molecular cloud contribute to the emission.

It is evident from Fig.~\ref{fig:GMCtot} that the radio ($\lambda \ga 0.1$mm) and the soft gamma ray ($\approx 1$ MeV $\div 1$ GeV) emission from the cloud is constant in time. This reflects the fact that such emission is produced by background CRs that enter the cloud. The emission in the other energy bands is variable in time, being produced by the CRs coming from the SNR. The flux of this latter component changes with time as indicated in Fig.~\ref{fig:crs}. The two most prominent features in the variable emission are two peaks, in X-- and gamma--rays respectively. 

The peak observable at gamma ray energies is the result of the decay of neutral pions produced when CRs of different energies coming from the SNR reach the cloud at different times. The peak moves at lower and lower energies with time, reflecting the fact that CRs with lower and lower energies progressively reach the cloud as time flows.
At early times, the emission can extend up to $\sim$ 100 TeV (solid line), revealing the presence of PeV CRs and thus indirectly the fact that the nearby SNR is acting as a CR pevatron \citep[see][for a discussion of this issue]{gabici2}.
Moreover, the gamma ray emission in the TeV range is enhanced with respect to the one expected from an isolated molecular cloud (dashed--dotted line in Fig.~\ref{fig:GMCtot}) for a period of several $10^4$ yr. This is much longer than the period during which SNRs are effectively accelerating the multi-TeV CRs responsible for the TeV emission, which lasts few thousands years. This is because the duration of the gamma ray emission from the cloud is determined by the time of propagation of CRs from the SNR to the cloud and not by the much shorter CR confinement time in the SNR. Therefore, the gamma ray emission from the cloud lasts much longer than the emission from the SNR, making the detection of clouds more probable \citep{gabici2}.

The peak in the X-ray spectrum is due to synchrotron emission from secondary electrons produced in CR interactions in the cloud. The peak is moving to lower energies with time but, unlike the gamma ray peak, it is also becoming less and less pronounced. This fact can be understood with the help of Fig.~\ref{fig:timeGMC} (left panel). After 500 yr from the supernova explosion (solid line in Fig.~\ref{fig:GMCtot}), PeV CRs from the SNR reach the cloud and produce there secondary electrons with energy in the $\approx 100$ TeV range. For these electrons, the synchrotron cooling time is comparable with the escaping time from the cloud. Thus, they release a considerable fraction of their energy in form of X-ray synchrotron photons before leaving the cloud. As time passes, lower energies CRs reach the cloud and secondary electrons with lower energies are produced. For these electrons the cooling time becomes progressively longer than the escape time and this explain the suppression of the synchrotron emission.
The X-ray synchrotron emission is weaker than the TeV emission for any time and for times $> 2000$ yr the ratio between TeV and keV emission can reach extreme values of a few tens or more.
These values are observed from some of the unidentified TeV sources such as HESS J1616-508 \citep{suzaku1, suzaku2} and more in general unidentified TeV sources are characterized by the absence of any clear counterpart at other wavelengths.
Due to this peculiar spectral properties, such sources have been labeled as {\it dark}, since they seemed to emit gamma rays only.
However, in the scenario presented in this paper, spectra showing a high TeV/kev flux ratio can be produced very naturally if a cloud is illuminated by CRs coming from a nearby SNR.
This suggestion is also supported by the fact that most of the unidentified TeV sources are spatially extended, as molecular clouds are expected to be.

In the hard X-ray/soft gamma ray region of the spectrum (from tens of keVs to hundreds of MeVs), partially covered by the INTEGRAL satellite, extremely hard spectra ($dN/dE \propto E^{-1}$) may result due to the Bremsstrahlung emission from primary and secondary CR electrons.

\begin{figure*}  
\resizebox{\textwidth}{!}{
\includegraphics{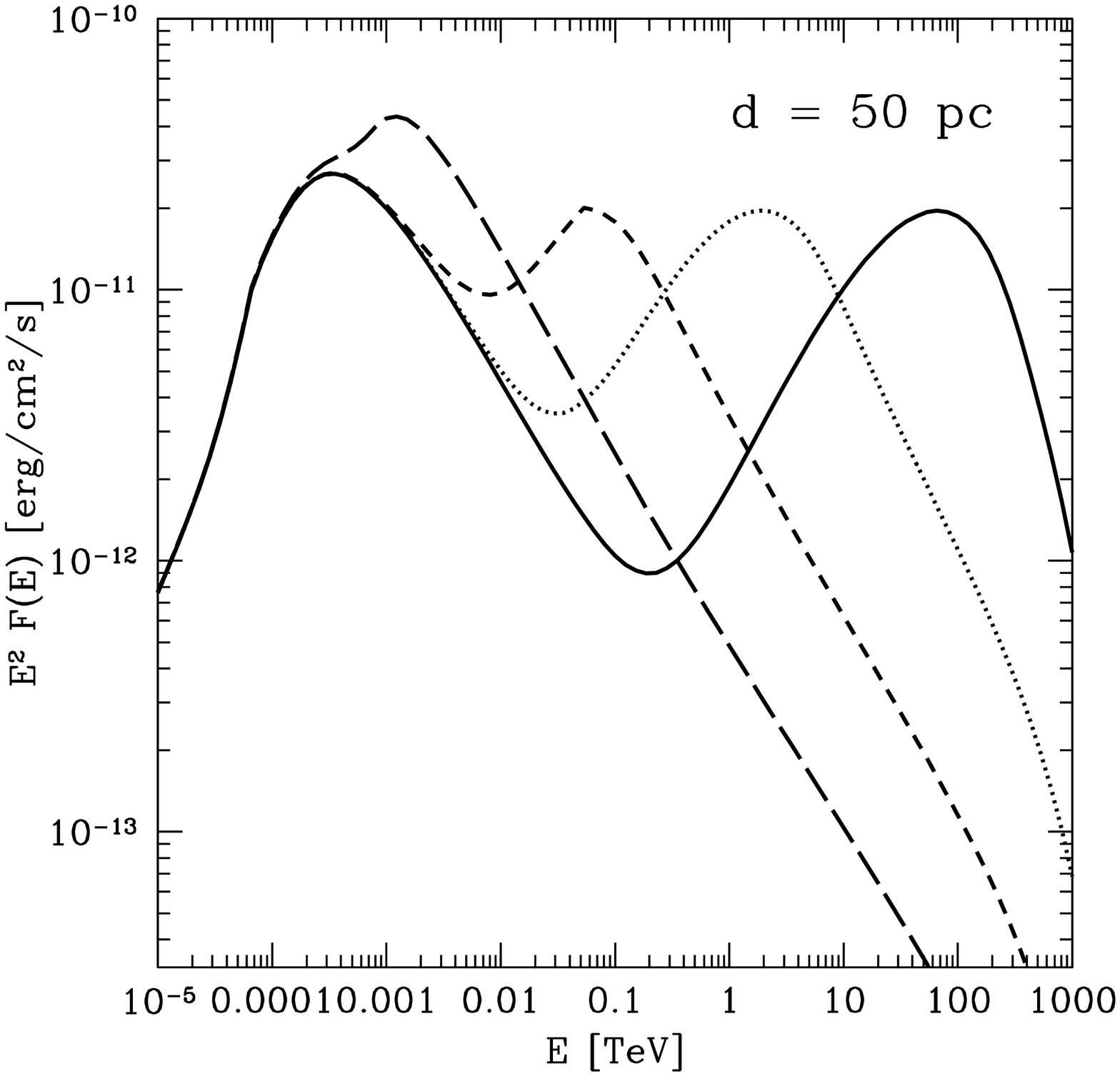}
\includegraphics{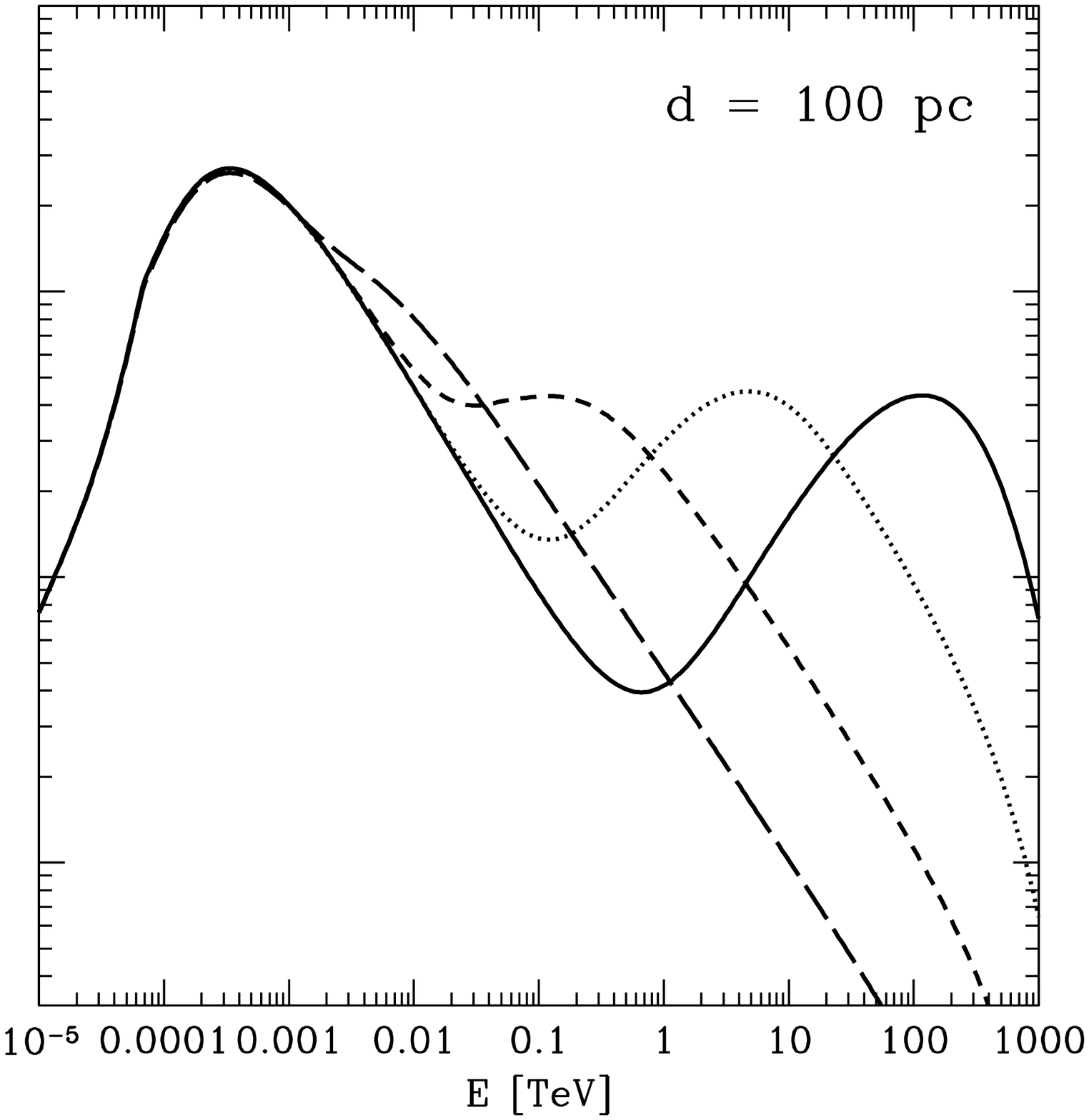}
\includegraphics{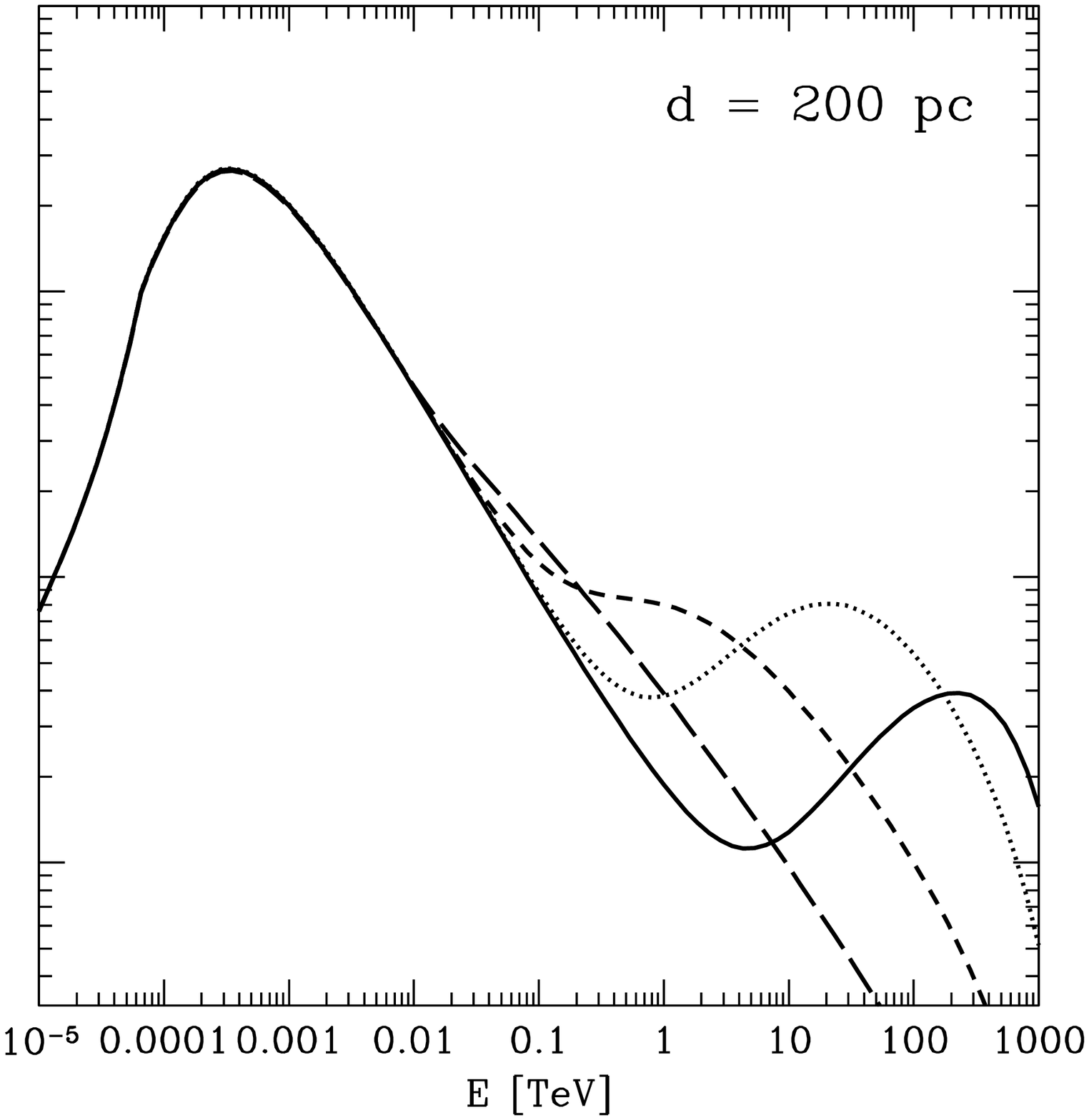}
} 
\caption{Total gamma ray emission from a molecular cloud of mass $10^5 M_{\odot}$ located at a distance of 1 kpc. The distance between the molecular cloud and the SNR is 50, 100 and 200 pc for left, centre and right panel, respectively. As in Fig.~\ref{fig:GMCtot}, the solid, dotted, short--dashed and long--dashed lines refers to the emission at a time 500, 2000, 8000 and 32000 years after the explosion.}
\label{fig:Vshaped}
\end{figure*}

Finally, the radio emission from the cloud is the result of the synchrotron emission of background CR electrons that penetrate the cloud. 
The contribution from secondary electrons is subdominant, due to the fact that the $\sim$ GeV secondary electrons that might emit synchrotron radio waves escape from the cloud before losing energy (see Fig.~\ref{fig:timeGMC}). 
Moreover, in the GeV energy range, ionization and Bremsstrahlung losses dominate, and this would further reduce the expected synchrotron emission.  
Recently, Protheroe et al. (2008) estimated the expected synchrotron radio emission from secondary electrons produced by CR interactions in the Sgr B2 giant molecular cloud.
They did not consider the contribution from background CR electrons and neglected the effects of diffusive transport of secondary electrons (namely, they assumed no penetration of electrons from outside the cloud and instantaneous cooling of secondary electrons produced inside the cloud).
These assumptions are valid for the very strong magnetic field of the order of one milliGauss measured for SgrB2. Such field is much  stronger than the one assumed here.
Thus, due to these somewhat extreme assumptions, possibly justified in the particular case of Sgr~B2 cloud, a direct comparison between their findings and the results presented in this paper is not straightforward.
Protheroe et al. (2008) also noticed that  the radio emission from a cloud at frequencies above $\sim 1$ GHz can be dominated by free--free thermal emission.
Being focused on the non-thermal emission from clouds, we did not attempt here to model their thermal emision. However, our findings can still be tested and constrained by observations in the GHz range.
This can be done by requiring the predicted synchrotron emission not to overproduce the observed free--free thermal emission.
A similar approach has been adopted by Jones et al. (2008).

\subsection{The spectral shape at GeV-TeV energies}

As noticed in the previous section, concave gamma ray spectra may be produced in a molecular cloud located in proximity of a SNR, as the result of the decay of neutral pions produced in CR interactions. Such concavity reflects the shape of the underlying CR spectrum, which consist of the superposition of two components: the galactic CR background, characterized by a steep spectrum, and the CRs coming from the nearby SNR, which exhibit a hard spectrum.
With this respect, the distance between the SNR and the molecular cloud $d_{cl}$ plays a crucial role.
This is because, the larger the distance between the SNR and the cloud, the lower the level of the CR flux coming from the SNR.
Moreover, also the time evolution of the emission from a cloud changes with $d_{cl}$ since the time it takes a particle with given energy to cover such a distance scales as $t \sim d_{cl}^2/D$, where $D$ is the diffusion coefficient.

In Fig.~\ref{fig:Vshaped} the total gamma ray spectrum from a molecular cloud is shown as a function of the distance between the SNR and the cloud. The cloud mass is $10^5 M_{\odot}$ and the distance from the SNR is 50, 100 and 200 pc for the left, central and right panel, respectively.
Similarly to Fig.~\ref{fig:GMCtot}, the solid, dotted, short-dashed and long-dashed lines refer to the emission for 500, 2000, 8000 and 32000 years after the supernova explosion.
It is evident from Fig.~\ref{fig:Vshaped} that a great variety of gamma ray spectra can be produced.
In almost the entirety of the cases considered, the gamma ray emission is characterized by the presence of two pronounced peaks.
The low energy peak, located in the GeV domain is steady in time and it is the result of the decay of neutral pions produced in hadronic interactions of background CRs in the dense intercloud gas. The high energy peak is the result of hadronic interactions of CRs coming from the nearby SNR, and thus it is moving in time to lower and lower energies (see previous section for a discussion).
Both the relative intensity and position of the two peaks depend on the distance between the SNR and the cloud.
Interestingly, the GeV emission from the cloud is affected by the presence of the nearby SNR only at late times after the explosion and only if the distance from the SNR is comparable or smaller than $\approx$ 50 pc (see Fig.~\ref{fig:Vshaped}, left panel).
In all the other cases the GeV emission is always the result of the interactions of background CRs and thus, at least in this case, observations of molecular clouds in the GeV gamma ray domain cannot be used to infer the presence of a CR accelerator located at a distance greater that $\approx$ 50 pc from the cloud. 

Similar concave or "V--shaped" spectra have been recently obtained by Rodriguez Marrero et al. (2008) in a different context in which two molecular clouds are assumed to be located in the proximity of a CR accelerator. If the two clouds happen to be located at different distances from the CR accelerator but within an angular separation smaller than the FERMI angular resolution, then they would appear as a single GeV source, and the superposition of their emission might result in concave spectra.
However, Rodriguez Marrero et al. (2008) did not include in their calculations the contribution to the emission coming from the ubiquitous galactic CR background, which in most cases dominates over the contribution of CRs from the nearby SNR,  at least for what concerns the GeV emission from the cloud (see e.g. Fig.~\ref{fig:Vshaped}).
Moreover, they also considered very short distances between the cloud and the CR accelerator (down to 5 pc), but in this case an accurate modeling  of the CR accelerator itself is needed, especially for what concerns its own gamma ray emission that might add up to the total emission.
In addition to that, distances as short as $\sim 5 \div 10$~pc are, under many circumstances, smaller than both the source size and the radius of the molecular cloud itself, and this changes significantly the problem since the interactions between the accelerator and the MC are likely to play an important role.

The prediction of V-shaped spectra that we make in this paper is more general than the one by Rodriguez Marrero et al. (2008), since it represents an intrinsic feature of a single molecular cloud which is located close to a CR source. The V-shaped gamma ray spectrum reflects the shape of the underlying CR spectrum which is the superposition of the steep spectrum of the background CRs and the hard spectrum of CRs coming from the nearby SNR.

\subsection{The role of the magnetic field}

The value of the magnetic field in the cloud is an important parameter since it regulates the synchrotron energy losses of high energy electrons, and the diffusive escape time of relativistic particles from the cloud.
Observations suggest that the value of the magnetic field in a molecular cloud scales with the square root of the gas density, reaching very high values (1 mG or more) in the very dense sub-parsec scale cloud cores \citep{crutcher}.
However, dense cores constitute a very small fraction of the total volume of molecular clouds.
Since here we are interested in calculating the emission from the whole cloud, the relevant parameter is the volume averaged value of the cloud magnetic field.
It seems reasonable to assume that the total magnetic energy in the cloud $W_B = (B^2/8 \pi) \times V$, where $V$ is the cloud volume, does not exceed the total gravitational energy of the cloud $W_G = 3 G M^2 / 5 R_{cl}$. 
This leads to a maximum value of the average magnetic field of: $B \le 30 ~ (M/10^5 M_{\odot}) (R_{cl}/20 {\rm pc})^{-2} \mu$G. 
This is in general agreement with observations, from which a value of $\approx 10 ~ \mu$G can be inferred for typical cloud densities of $\approx 100$~cm$^{-3}$.
However, since the dispersion around this mean value is considerable \citep[see e.g.][]{crutcher}, it is worth to investigate how the non thermal emission from a molecular cloud depends on the actual value of the magnetic field.

\begin{figure}  
\resizebox{.5\textwidth}{!}{
\includegraphics{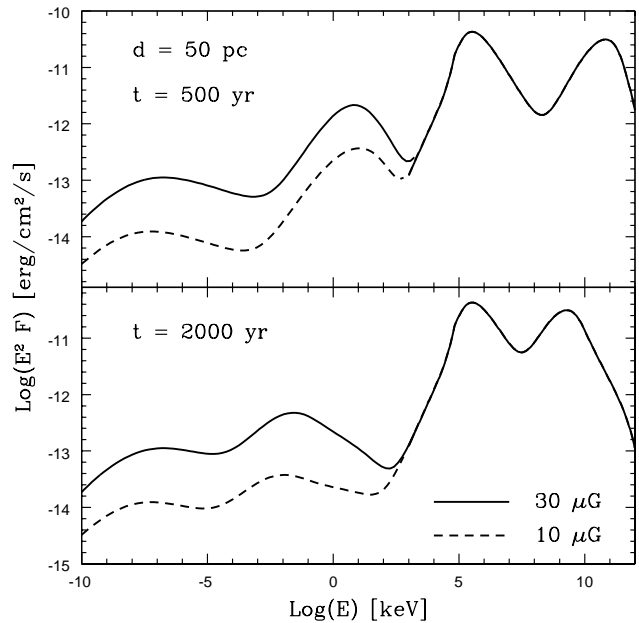}
} 
\caption{Broad band emission from a molecular cloud of mass $10^5 M_{\odot}$, radius $20$~pc, density $\sim 120$~cm$^{-3}$. The distance between the cloud and the SNR is 50 pc. Top and bottom panels refer to 500  and 2000 years after the supernova explosion, respectively. The magnetic field is 30 $\mu$G (solid lines) and 10 $\mu$G (dashed lines).}
\label{fig:Bfield}
\end{figure}

Fig.~\ref{fig:Bfield} shows the broad band spectrum from a molecular cloud of mass $10^5 M_{\odot}$, radius $20$~pc, density $\sim 120$~cm$^{-3}$. The distance between the cloud and the SNR is 50 pc. 
In the top and bottom panels is plotted the cloud emission after 500 and 2000 years from the supernova explosion, respectively.
The solid line refers to a value of the magnetic field of 30 $\mu$G, while the dashed line represents the emission for a smaller value of the field equal to 10 $\mu$G.
The cloud emission from radio frequencies up to the hard X-ray band strongly depends on the value of the magnetic field, while the gamma ray emission is unaffected, being the result of hadronic interactions of CR protons.
The strong dependence of the radio and X-ray emission on the magnetic fields is evident, and this demonstrate that this effect has to be taken into account in multi wavelength studies of molecular clouds.

\section{Conclusions}

The non thermal emission from a molecular cloud located in proximity of a SNR is the result of the interactions of CRs that penetrate the cloud.
Both CRs in the galactic background and runaway CRs from the SNR contribute to the emission.
In this paper, we calculated the expected non thermal emission from the molecular cloud as a function of several parameters, such as the time after the supernova explosion, the distance between SNR and cloud, and the cloud magnetic field.
In all the cases, the gamma ray emission from the cloud is by far exceeding the energy output in other energy bands.

The gamma ray emission from the cloud consists of two distinct components.
The first one is due to the interactions of CRs from the galactic background, which are characterized by a steep spectrum.
This component is steady in time and peaks in the GeV energy region.
The second component is the result of the interactions of runaway CRs that escape the SNR and diffusively reach the cloud.
The spectrum of these runaway CRs is hard and variable in time and both this characteristics are reflected in the related gamma ray emission.
In particular, this component is producing a second peak in the gamma ray spectrum which moves to lower and lower energies, reflecting the fact that high energy CRs are released from the SNR earlier and diffuse faster than low energy ones, and as a consequence they reach the cloud earlier.
The superposition of these two components of the gamma ray emission may produce concave spectra, which are steep at low ($\sim$ GeV) energies and hard at high ($\sim$ TeV) energies.
This peculiar spectra might be revealed by means of joint observations of FERMI and ground based Cherenkov telescopes.
The detection of such emission would provide a strong hint for the presence of a CR accelerator in the neighbourhood of the molecular cloud.

It has been suggested that some of the {\it dark} TeV unidentified sources, with no obvious counterparts at any other wavelength, might be indeed associated with molecular clouds illuminated by CRs from nearby accelerators such as SNRs \citep{atoyan,gabici2}.
We showed here that, in this scenario, the {\it "darkness"} of a source, often defined as the radio $R_{TeV/X}$ between the TeV and X-ray observed flux, depends on several parameters such as the distance between the SNR and the cloud, the time after the supernova explosion and the value of the magnetic field in the molecular cloud.
However, in most cases it is very natural to obtain very high values for $R_{TeV/X}$, compatible with the estimates on $R_{TeV/X}$ claimed for some of the unidentified TeV sources \citep[see e.g.][]{suzaku2,suzaku3}.
This further supports an association between unidentified TeV sources and molecular clouds and suggests that the unknown nature of these objects might be soon revealed.

\label{lastpage}

\end{document}